\documentclass[12pt]{article}

\usepackage{amssymb}
\usepackage{epsfig}
\usepackage{graphicx,color}
\usepackage{amsmath}

\newcommand{\beqa}{\begin{eqnarray}}
\newcommand{\eeqa}{\end{eqnarray}}

\begin{document}
\def\ii{\'\i}

\title{
The concept of nuclear cluster forbiddenness
}

\author{
H. Y\'epez-Mart\ii nez$^1$, P. O. Hess$^2$
}

\author{H. Y\'epez Mart\ii nez$^1$, P. O. Hess$^2$ \\
{\small\it
$^1$Universidad Aut\'onoma de la Ciudad de M\'exico,
Prolongaci\'on San Isidro 151,} \\
{\small\it
Col. San Lorenzo Tezonco, Del. Iztapalapa,
09790 M\'exico D.F., Mexico} \\
{\small\it
$^2$Instituto de Ciencias Nucleares, UNAM, Circuito Exterior, C.U.,}\\ 
{\small\it A.P. 70-543, 04510 M\'exico D.F., Mexico} }

\maketitle

\abstract{
In nuclear cluster systems, a rigorous structural forbiddenness of virtual nuclear division into unexcited fragments is obtained.  We re-analyze the concept of {\it forbiddenness}, introduced in \cite{smirnov} for the understanding of structural effects in nuclear cluster physics. We show that the concept is more involved than the one presented previously, where some  errors were committed. 
Due to its importance, it is reanalyzed here.
In 
the present contribution a simple way for the determination of forbiddenness
is given, which may easily be extended to any number of clusters, though in this contribution we discuss only two-cluster systems, 
for illustrative reasons.
A simple 
rule 
is obtained for the minimization of the forbiddenness, namely to start from 
a cluster system with a large $SU(3)$ irrep
$\left(\lambda_c,\mu_c\right)$, but minimizing
$\left(\lambda_c-\mu_c\right)$, i.e. the system has to be
oblate.  
The rule can be easily implemented in structural studies, done up to now with an oversimplified definition
of {\it forbiddenness}.
The new method is applied to various systems of light clusters and to some decay channels of $^{236}$U
and $^{252}$Cf.  
}

\section{Introduction}
\label{intro}

In a system of two light clusters,
each nucleus can  be well described within
the $SU(3)$ model of the nucleus.  The relation between the shell and the cluster model was established
by Wildermuth and Kanellopoulos \cite{wildermuth-0}, 
proving that within the harmonic oscillator the Hamiltonians of these two models can be related to each other
exactly.
 For
heavy clusters the pseudo-$SU(3)$ \cite{hecht,arima}
has to be applied, which
takes into account effectively the spin-orbit interaction. 
The \,\, $SU(3)$ model is based on the harmonic oscillator as a mean field, plus residual interactions.
In \cite{wildermuth} a minimal condition is introduced for satisfying 
the Pauli exclusion 
principle: When the sum of the number of oscillation quanta of each cluster
is determined, obtained by filling densely the nucleons into the oscillator states,
a mismatch is obtained to the total number of oscillation quanta in the united
nucleus. In order to observe minimally the Pauli exclusion principle, the
missing oscillation quanta ($n_0$) are usually included in the relative motion of
the two clusters. This constraint is known as the {\it Wildermuth condition}.
In \cite{sacm1,sacm2} additional requirements are added, such that the 
final $SU(3)$
irreducible representations (irrep) also fulfill the Pauli exclusion principle.
In \cite{sacm1,sacm2} light clusters were considered and the external product of
the cluster $SU(3)$ irreps with the irrep of 
the relative motion always can be coupled to the
ground state and excited states of the united nucleus, i.e., it suffices to add the missing $n_0$ quanta to the relative motion only.

It was proven in \cite{smirnov} that it is not always the case
that the relative $SU(3)$ irrep, required by the 
{\it Wildermuth condition} can be 
coupled with the cluster irrep, $(\lambda_c,\mu_c)$ to the ground state
of the united nucleus. A rigorous structural {\it forbiddenness}
(in the sense that certain combinations of representations
are forbidden)
of virtual nuclear division into two unexcited fragments exists for all
nuclei   with
masses $A_1\,,\ A_2  > 36$, and at  $A_1\,,\ A_2  > 12$, for a wide range of nuclei \cite{smirnov}. The {\it forbiddenness} is universal and model independent. 

In the literature there are known two apparently 
different definitions
for cluster states:
i) One definition requires a rigid molecular type separation while ii) others
(including ourselves) speak both about localized and shell-model-like clusters. For a detailed review related to the first definition, see the report given in \cite{horiuchi}. 
The second one
is closely related to experimental observability: A state is called a cluster state when its wave function has an overlap with cluster wave functions \cite{o16}, 
as for example in the case
of $^{16}$O $\rightarrow$ $^{12}$C+$\alpha$. This
definition is more general and includes the rigid definition
of cluster states \cite{darai-2012}. 
A detailed discussion is given in \cite{RMF2008}.

In order to understand basic processes, as fission and fusion or
phenomena as cluster radioactivity 
\cite{sandulescu,poenaru,gupta}, 
structural considerations play a very important role
modern nuclear physics. For example, in \cite{algora-2006}
preferences of clusterizations in ternary fission processes
were investigated. Hyperdeformed states in $^{36}$Ar were
considered in \cite{cseh2010}, compared to experiments, and in \cite{darai-2011} the clusterization in shape isomers of $^{58}$Ni. In all cases structural considerations lead to a fundamental understanding of these systems, which is of course not sufficient and energetic considerations 
\cite{algora-2006,darai-2012} are necessary as well as
the penetration through the barrier
\cite{spec-fac-serban}.

Structural considerations play a role in understanding
the connection between 
fission and cluster radioactivity, often seen
as different processes. In \cite{spec-fac-serban} 
it was argued that they
are the same, using the pseudo-$SU(3)$ model
and determining the spectroscopic factors within an algebraic
cluster model. Other applications are related to the understanding of structural aspects in radiation capture
\cite{cseh-2012}.

Most of these theoretical considerations are based on the
shell model, which has been and still is
the {\it cornerstone} of microscopic nuclear structure.
Although it can be argued that some nuclear states
as the Hoyle state
\cite{hoyle} can not be described by a single $SU(3)$ irreducible representation
\cite{draayer-hoyle} or that $SU(3)$ is strongly broken for heavy
nuclei, giving the impression that $SU(3)$ is out of date, however this is not true.
For heavy nuclei methods were developed as the pseudo-$SU(3)$
\cite{hecht,arima} or the use of embedded (effective)
$SU(3)$ representations 
\cite{embedded1,embedded2,embedded3}, which was finally 
applied with great success in \cite{hunyadi}. In all these
extensions the language of the harmonic oscillator ($SU(3)$)
was applied  \cite{o16}, which proved to be very successful for
light nuclei and served as a starting point for
different approximations for heavy nuclei.
The considerations made in this contribution can be applied
to all these extensions, with slight modifications.
The $SU(3)$ language
allows to understand many of features in  modern nuclear physics experiments without recurring to complicated numerical
routines, though ab-inito calculations  
\cite{horiuchi,draayer-hoyle} 
may be
necessary to understand the details.
The report on modern nuclear physics \cite{horiuchi} 
concentrates for a large part on the 
$SU(3)$ notation and simple structural considerations, 
showing that the harmonic oscillator helps to understand 
up-initio calculations.
The theory, presented in \cite{draayer-hoyle}, also used primarily the harmonic shell-model
picture, including inter-shell excitations. The calculation is restricted to only a few
$SU(3)$ irreps as band-heads of symplectic excitations,
which start from $SU(3)$ bad heads in the valence shell
and adds $2\hbar\omega$ excitations. This model has a lot in common with
the {\it Semimicroscopic Alegbraic Cluster Model} (SACM) \cite{sacm1,sacm2}, which also starts from
$SU(3)$ irreps in the 0$\hbar\omega$ shell and includes intershell excitations by adding
relative oscillation quanta. Due to its starting point of a cluster picture, it is able
to describe complicated cluster configurations. 

In \cite{smirnov} it is observed that the coupling in $SU(3)$ of two light clusters, in their ground state, 
with the relative motion not always leads
to the ground state of the united nucleus. Thus, some structural
information is missing. Due to that,
in \cite{smirnov} the concept of {\it  forbiddenness } was introduced.
The main idea is to allow the excitation of one or both clusters to higher shells,
subtracting the number of excitation quanta from the relative motion.
This increases the possibility to reach the ground state of the united nucleus.
The minimal
quanta to excite one of the clusters, needed to achieve this 
goal, is called the  {\it  forbiddenness}. This property was interpreted in Ref. \cite{Yu}
and, independently, by Bader and Kramer \cite{Bader} as a
consequence of the $SU(3)$ symmetry selection rules.  In \cite{smirnov}, 
a simple example
was discussed in order to illustrate the 
calculation of the  {\it forbiddenness},  though the system
itself does not play an important role today:
The analysis of the forbidden
decay $A \longrightarrow A_1 +A_2$,
for the case of
two spherical clusters ($^{40}$Ca) and 
a spherical united nucleus $^{80}Zr$, 
which is simple enough to understand the concept
of forbiddenness.  
Smirnov et al. \cite{smirnov} 
obtained for  the above cluster system the {\it forbiddenness}  value of 20.
However, we were unable to reproduce the results obtained in \cite{smirnov}, which leads to the present contribution
and in what follows we resume the prove why in 
\cite{smirnov} an error must have been committed: 
According to the Wildermuth condition, the minimal 
number of quanta to add 
is $n_0=60$, which is distributed 
between the relative oscillation quanta and the
excitation of the clusters. When the number of relative quanta is $n_r$, the irrep
of the cluster state {\it has to be} $(0,n_r)$ in order 
to couple to a total scalar irrep
(0,0) of $^{80}$Zr.  The Young diagram for the irrep $(0,n_r)$ of the
coupled cluster $^{40}$Ca+$^{40}$Ca has $2n_r$ intershell excitations and is given 
by $\left[40+n_r,40+n_r,40\right]$. From this, it is clear that adding the relative
motion irrep $(n_r,0)$, the total number is $3n_r=n_0=60$ and
the number of  $n_c\hbar\omega$\, excitation in the cluster irrep has to be 
$2n_r=40$, not 20 as reported previously in \cite{smirnov}. 
The above result demand a reconsideration
of the concept of forbiddenness. Due to that,
the presentation will tend to be more mathematical but the
simple results, 
easy to implement in recent cluster calculations
( see \cite{algora-2006,cseh2010,darai-2011,cseh-forbiddeness,cseh-tri-cluster}
and references therein)
and their importance for current studies in
nuclear physics justifies it. 

Due to the importance of the concept of {\it forbiddenness}
in modern nuclear structural analysis and the impossibility to
reproduce the results in \cite{smirnov} an alternative definition was proposed in
\cite{cseh-forbiddeness}, which simply depends on the 
difference
in the $SU(3)$ irrep of the ground state of the 
united nucleus to the ones 
in the final coupling of the cluster system. This new
definition was used in 
\cite{darai-2011,cseh-2012,cseh-2009}, proving its
utility.

We consider it very
important to present a consistent manner to determine 
easily the {\it forbiddenness},
which explains, according to \cite{smirnov} why certain cluster combinations
are suppressed in fission and/or fusion processes. {\it The conclusions 
remain the same}: Structural considerations, coming solely from the
coupling of nucleons to $SU(3)$ irreps, are sufficient in order to understand
the suppression and enhancements of cluster distributions in nuclear reactions.
Of course, for the details one has to superimpose tunnel effects \cite{spec-fac-serban}. 
This conclusion was also obtained in 
\cite{cseh-forbiddeness,cseh-tri-cluster}.
The concept of forbiddennes, as will be exposed in this contribution, can
be applied to heavy nuclei too, using the extensions of the
harmonic oscillator picture, as mentioned above. It has an important impact on super-heavy
nuclei \cite{serban-heavy}.

Due to the reasons explained above one has to reanalyze the
concept of {\it forbiddenness}, which requires some group theory.
Therefore, within the scope of this contribution we 
will concentrate on the
forbiddenness and only indicate its use.
It is out of the scope of this publication to do an explicit
investigation neither on fusion and fission processes nor
cluster radioactivity, though a couple of examples will
be discussed at the end, constraint to the determination of
the {\it forbiddenness}.

\section{A new proposal for forbiddenness}

In what follows, we will present a detailed corrected derivation of 
the {\it forbiddenness} 
and demonstrate that it depends on the {\it compactness}  
of the two clusters,
before the relative motion is added. The  
{\it compactness} is defined by
the eigenvalue of the 
second order Casimir operator of $SU_c(3)$, with 
respect to the irrep $(\lambda_c,\mu_c)$, and 
the difference $\left(\lambda_c-\mu_c\right)$.
The larger the eigenvalue is but larger the difference of 
$\mu_c$ to $\lambda_c$ (oblate),
the less oscillation quanta have to be
invested in the excitation of the clusters. For the case of two
spherical clusters and a spherical united nucleus, 
where the only irrep in the $0\hbar\omega$ shell is (0,0),
the {\it forbiddeness} is identical to include only inter-shell excitations in the cluster system.
However, especially when both
clusters and the united nucleus are deformed, 
various values for 
the {\it compactness} are possible.

Let us first consider a two-cluster system, coupled to the 
{\it cluster irrep} $(\lambda_c,\mu_c)$.   This irrep  is obtained by 
coupling the ground state 
irreps $(\lambda_i,\mu_i)$ ($i=1,2$) of the two clusters to 
$(\lambda_c ,\mu_c)$.
In 
contrast to the former assumption that each cluster has to be in its ground state, each one is allowed 
to be excited, 
including $n\hbar\omega$ inter-shell excitations. We are not interested on how each
cluster is excited but refer to the resulting cluster irrep
as $(\lambda_c +\lambda _0,\mu_c+\mu _0)$, where $\lambda_0$ and $\mu_0$ are given in terms of the internal excitations of the two joined clusters. 
In what follows, an analytical expression for these internal excitations will be obtained. 

The counting procedure starts with the product of the 
{\it excited} cluster irrep  $(\lambda_c +\lambda _0,\mu_c+\mu _0)$  with
the relative motion $(n_r,0)$, in order to obtain the irrep of the parent nucleus $\left( \lambda , \mu \right)$. When 
the $SU(3)$ irreps are converted to Young diagrams,
the product reads 

\beqa
&
\left[ \lambda_c+\mu_c+a,\mu_c+b,c\right] \otimes \left[n_r,0,0\right]
&
\nonumber \\
&~\rightarrow~& 
\nonumber \\
&
 \left[ \lambda_c+\mu_c+a+k_1,\mu_c+b+k_2,c+k_3\right]
~~~,
&
\label{prod}
\eeqa
where

\beqa
&
n_0 ~=~ n_c+n_r ~,~ n_c=a+b+c ~,~ n_r ~=~ k_1+k_2+k_3
~~~,
&
\nonumber \\
&\lambda_0  =  a-b ~,~ \mu_0 ~=~ b-c &
\label{prod-0}
\eeqa
and $n_0$   is the minimal number of oscillation quanta to be added, in order 
to guarantee the Pauli 
exclusion principle. These quanta are divided in
the excitation $n_c$ of the cluster irrep and the remaining relative 
oscillation quanta, denoted by $n_r$. 
The {\it forbiddenness} is given by
the excitation $n_c$ of the cluster 
irrep and has to be as small as possible. We can start by setting
$c=0$. The final $SU(3)$ irrep is $\left( \lambda , \mu \right)$ and
represents the final product of Young diagrams
in (\ref{prod}).

Using the rules for the direct product of an arbitrary Young diagram with a 
symmetric one \cite{hamermesh}, we obtain the following 
additional equations and constraints:

\beqa
0 & \leq & k_2 ~\leq~ \lambda_c + \lambda_0 
\nonumber \\
0 & \leq & k_3 ~\leq~ \mu_c + \mu_0 
\nonumber \\
\lambda & = & \lambda_c + \lambda_0 + k_1-k_2
\nonumber \\
\mu & = & \mu_c + \mu_0 + k_2 - k_3
~~~.
\label{eq-1-b}
\eeqa ``$a$" represents the number of boxes which are added in the first row of the cluster irrep 
(Young diagram: $\left[ \lambda_c+\mu_c+a,\mu_c+b,0\right]$),  ``$b$" is the number of boxes which are added to the second row. 

From the third and fourth equation in (\ref{eq-1-b}) we get for $k_1$ and $k_3$

\beqa
k_1 & = & \lambda - \lambda_c - \lambda_0 + k_2
\nonumber \\
k_3 & = & \mu_c - \mu + k_2 + \mu_0
~~~,
\label{eq-3-b}
\eeqa
from which we obtain

\beqa
n_r & = & \left( \lambda - \lambda_c - \lambda_0 + k_2 \right) + k_2
+\left( \mu_c + \mu_0 - \mu + k_2 \right)
\nonumber \\
& = &
(\lambda - \mu ) -(\lambda_0 - \mu_0) - (\lambda_c-\mu_c)+3k_2
~~~.
\label{eq-4-b}
\eeqa

With this, we get for the total number of quanta 
which have to be added according to the Wildermuth condition:

\beqa
n_0 & = & n_r + n_c ~=~ n_r +(\lambda_0+2\mu_0)
\nonumber \\
& = & (\lambda - \mu )- (\lambda_c - \mu_c) +3(\mu_0+k_2)
~~~.
\label{eq-5}
\eeqa
Resolving for $(\mu_0+k_2)$, we get

\beqa
\mu_0 + k_2 & = &
\frac{1}{3} \left[
n_0+(\mu - \lambda ) -(\mu_c -\lambda_c)
\right]
~~~,
\label{eq-6}
\eeqa
which is a {\it fixed quantity}, because all values on the right hand side
of the equation are specified for a system considered.

The last equation is substituted into the second one in (\ref{eq-3-b}), giving

\beqa
k_3 & = & \frac{1}{3}\left[ n_0 -(\lambda +2\mu )+(\lambda_c+2\mu_c) \right]
~~~,
\label{eq-7}
\eeqa
{\it which is also a fixed quantity}!

Now, we use the first equation in (\ref{eq-3-b}), for $k_1$,
and also use (\ref{eq-6}), for $k_2$, yielding

\beqa
k_1 & = & \frac{1}{3} \left[
n_0 + (2\lambda +\mu )-(2\lambda_c+\mu_c)\right]
-(\lambda_0+\mu_0)
~~~.
\label{eq-8}
\eeqa

Let us summarize the results for the $k_i$ in a slight different form:

\beqa
k_1+\lambda_0+\mu_0  & = & \frac{1}{3} \left[
n_0 + (2\lambda +\mu )-(2\lambda_c+\mu_c)\right]
\nonumber \\
k_2+\mu_0 & = &
\frac{1}{3} \left[
n_0+(\mu - \lambda ) -(\mu_c -\lambda_c)
\right]
\nonumber \\
k_3 & = & \frac{1}{3}\left[ n_0 -(\lambda +2\mu )+(\lambda_c+2\mu_c) \right]
~~~.
\label{eq-9}
\eeqa

Using that $n_{r}=k_{1}+k_{2}+k_{3}$, $n_{c}=\lambda
_{0}+2\mu _{0}$ and substituting only $k_1$ and $k_3$, we get

\begin{equation}
n_{0}=n_{c}+n_{r}~=~\frac{1}{3}\left( 2n_{0}+\lambda -\mu -\lambda _{c}+\mu
_{c}\right) +k_{2}+\mu _{0}
~~~.  
\label{no-check}
\end{equation}

For $(\mu _{0}+k_{2})$, we use (\ref{eq-6}) and 
resolve (\ref{no-check}) for $n_c$, requiring that $n_c$ is minimal, giving

\beqa
n_{c} ~=~ n_0 - n_r & = & 
\overbrace{\left( n_{0}-k_{3}\right) }^{fixed}
-\overbrace{(k_{1}+k_{2})}^{\max}
~~~.
\label{nc-new}
\eeqa
We joint $n_0$ and $k_3$ because these values are already fixed.

The $n_{c}$ is minimized, when
$\overbrace{(k_{1}+k_{2})}^{\max}$ is maximized. Taking into account the 
first relation in (\ref{eq-1-b}) and the first equation in (\ref{eq-3-b}), 
according to which
$k_2^{{\rm max}}=\lambda _{c}+\lambda _{0}$ and $k_{1}^{\max }=\lambda +k_{2}^{\max }-\left(
\lambda _{c}+\lambda _{0}\right) =\lambda $
and using (\ref{eq-9}), we obtain

\beqa
\overbrace{(k_{1}+k_{2})}^{\max}~ &=&\lambda +\lambda _{c}+\lambda _{0}
\nonumber \\
&=&\left\{ \frac{1}{3}\left[ n_{0}+(2\lambda +\mu )-(2\lambda _{c}+\mu _{c})
\right] -(\lambda _{0}+\mu _{0})\right\} 
\nonumber \\
&&+\left\{ \frac{1}{3}\left[ n_{0}+(\mu -\lambda )-(\mu _{c}-\lambda _{c})
\right] -\mu _{0}\right\}  
\nonumber \\
& = & \frac{1}{3}\left[
2n_0+ \left( ´\lambda + 2\mu \right)-\left( ´\lambda_c + 2\mu_c \right)
\right]
-\left(\lambda_0+2\mu_0\right)
~~~.
\label{minimize1}
\eeqa
Resolving for $\left( \lambda_0 + \mu_0\right)$ = $a^{{\rm min}}$,
we get for 

\beqa
a^{{\rm min}} & = & {\rm max}\left[ 0,(\lambda _{0}+\mu _{0})\right]
\nonumber \\
& = &  {\rm max}\left[0,\frac{1}{3}\left\{ n_{0}-(\lambda -\mu
)-(2\lambda _{c}+\mu _{c})\right\}\right]
~~~,
\eeqa
having used $\lambda_0=a-b$ and $\mu_0 = b$. When the expression depending
on the irrep numbers is negative, one has to take the value zero.

Knowing $a^{{\rm min}}$,
minimizing $n_{c}=\lambda _{0}+2\mu _{0}=a+b$ has been now reduced to
minimize $\mu _{0}=b$. The $b^{{\rm min}}$ can be obtained considering that
(see (\ref{eq-1-b}))

\beqa
\frac{1}{3}\left[ n_{0}-(\lambda +2\mu )+(\lambda _{c}+2\mu _{c})\right]
&=&k_{3}~\leq ~\mu _{c}+\mu _{0} 
\eeqa
or
\beqa
\mu _{0} &\geq &-~\mu _{c}+\frac{1}{3}\left[ n_{0}-(\lambda +2\mu )+(\lambda
_{c}+2\mu _{c})\right]
~~~.
\eeqa
Because $\mu_0=b$ and only positive values are allowed, we obtain

\beqa
b^{{\rm min}} & = & 
{\rm max}\left[ 0,
\frac{1}{3}\left\{ n_{0}-(\lambda +2\mu )
+(\lambda _{c}-\mu_{c})\right\} \right]  
\label{mu-min}
~~~,
\eeqa
where again one considers 
the possibility that the expression in terms
of $n_0$ and the $SU(3)$ irreps may be negative. In this case the value 
$b^{{\rm min}}=0$ has
to be chosen. 

With this, we have minimized the {\it forbiddenness}

\beqa
n_{c} & = & 
\left( \lambda _{0}+2\mu _{0}\right) ^{\min }=\left( a\right) ^{\min}+\left( b\right) ^{\min}
\nonumber 
\\
& = & 
{\rm max}\left[0,\frac{1}{3}\left\{ n_{0}-(\lambda -\mu
)-(2\lambda _{c}+\mu _{c})\right\} \right] 
\nonumber \\
&& +{\rm max}\left[ 0, \frac{1}{3}\left\{ n_{0}-(\lambda +2\mu )
+(\lambda _{c}-\mu_{c})\right\} \right]
~~~.
\label{forbid-min}
\eeqa

Eq. (\ref{forbid-min}) 
is the main result of this contribution and
can be interpreted as follows: 
The first term in (\ref{forbid-min}) tells us, that 
in order to {\it minimize} $n_c$, we
have to {\it maximize} $(\lambda_c+2\mu_c)$.
The second term tells us that in addition the difference
$\left(\lambda_c - \mu_c\right)$ has to be minimized. A maximal 
$(\lambda_c+2\mu_c)$ and a minimal $\left(\lambda_c - \mu_c\right)$
imply a large compact and oblate 
configuration of the two-cluster system.

One can achieve these conditions, now defined as 
{\it compactness} of the irrep 
$\left(\lambda , \mu\right)$, determining the whole product
of $(\lambda_1,\mu_1) \otimes (\lambda_2,\mu_2)$ 
and searching for the irrep that corresponds to a large
compact structure (large $\left(2\lambda_c+\mu_c\right)$ but with a maximal difference $\left(\mu_c-\lambda_c\right)$, 
leading to an oblate structure).
For deformed clusters, there is also 
the possibility to excite it within the $0\hbar \omega$, leading to other 
individual cluster irreps $(\lambda_i,\mu_i)$. 
One can take the whole 0$\hbar\omega$ space of
each cluster and multiply them all, or even one can do it for the proton space and
the neutron space for each cluster and then multiply the proton final space with
the neutron final space. 

In agreement
to Smirnov's definition, \, the {\it forbiddenness} is related to 
$n_c \hbar\omega$  {\it excitations}  \, of a cluster system before 
adding the relative oscillation quanta. {\it But}, 
the excitation might be distributed
between the clusters and is a combination of these cluster excitations in 
0$\hbar\omega$ and excitation of these clusters in $n_c\hbar\omega$, with $n_c>0$.

With the derivation given above, we can address the example discussed in \cite{smirnov} and in the introduction,
namely $^{40}$Ca+$^{40}$Ca $\rightarrow$ $^{80}$ Zr. We have $(0,0)$ for both $^{40}$Ca-clusters, thus, $(\lambda_c,\mu_c)=(0,0)$ 
is the only possibility.
The irrep for  $^{80}$Zr is $(0,0)$ and $n_0=60$ according to the
Wildermuth condition, thus
$n_c  =  \frac{2n_0}{3} ~=~ 40$.
There is no higher $0\hbar\omega$ cluster irrep  $(\lambda_c,\mu_c)$, \, because each cluster has 
(0,0), also when we consider protons and neutrons separately. Thus, in this
example the solution is  unique.  As noted above, in \cite{smirnov} a 
forbiddenness of 20 is reported, which is incorrect.

The next, non-trivial example is $^{20}$Ne+$^{16}$O$\rightarrow$ $^{36}$Ar, when 
the clusters are in their ground state, we obtain the following list of the irreps
for the cluster and total irrep:

\beqa
&
\left( \lambda ,\mu \right) _{_{18}^{36}Ar_{18}} ~=~ (0,8)  ~,~
\left( \lambda _{1},\mu _{1}\right) _{_{8}^{16}O_{8}} ~=~ (0,0)  
&
\nonumber \\
&
\left( \lambda _{2},\mu _{2}\right) _{_{10}^{20}Ne_{10}} ~=~ (8,0)  ~,~
\left( \lambda _{c},\mu _{c}\right) ~=~ (8,0) ~,~
\left( n_{0},0\right) ~=~ (20,0) 
~~~.
&
\label{eq-19}
\eeqa
Taking the product of $(\lambda_c,\mu_c)$ with $(n_c,0)$ the resulting irreps
have the structure $(\lambda_c+a-b,\mu_c+b)$.  Using the 
equation (\ref{forbid-min}), 
we obtain for $a=\lambda_0+\mu_0=4$, $b=\mu_0=4$ and for the {\it forbiddeness} 
$n_c=8$. The $k_i$ values are: \, $k_1=0$, $k_2=8$ and $k_3=4$.\, 

A further, trivial example is $^{12}$C+ $^{8}$ Be $\rightarrow$ $^{20}$ Ne, 
whose $SU(3)$ irreps are $\left( \lambda _{1},\mu _{1}\right) _{^{12}C} =(0,4)$, $\left( \lambda _{2},\mu _{2}\right) _{^{8}Be} =(4,0)$ and $\left( n_{0},0\right)=(8,0)$, respectively.\,
In this case, $a^{{\rm min}}$ and $b^{{\rm min}}$ are both zero, thus,
the {\it forbiddenness}  \,is zero too, which is a known result.

In what follows, 
we present the results for 
$^{236}$U, decaying into two clusters as it happens
in a fission process. 
Many of the clusters involved are not in the p or sd-shell, thus the
$SU(3)$ model of Elliott \cite{elliott} does not work due to the large spin-orbit
interaction. One has to apply the pseudo-$SU(3)$ model \cite{hecht,arima}.
This model takes into account the spin-orbit interaction by renaming the
spin and orbital angular momentum to pseudo-spin and pseudo-angular momentum.
The space is divided into {\it active} nucleons and {\it spectators}. 
The spectators are all nucleons in the intruder orbital levels $j+\frac{1}{2}$,
while the active nucleons are in the remaining orbitals.
It would
lead outside the scope of this presentation to give an explicit description
of this model but rather refer to applications, like in 
\cite{serban-heavy,pseudo-symplectic}. These references show in detail how to determine the
pseudo-$SU(3)$ irreps for each cluster and the united 
nucleus. The coupling of 
the cluster irreps with the relative motion and all considerations made above stay the same. 

Here, we will resume the main steps and present the results in Tables \ref{Tx1}
and \ref{Ty2}:
In the system of the united nucleus the nucleons are filled into the Nilsson
levels at its deformation
$\epsilon_2 = 0.2$, taken from \cite{moller}. The 
nucleons in the active orbitals are filled into
the levels of the pseudo-oscillator, which provides the pseudo-$SU(3)$ irrep,
in the same way as has been done for light nuclei within the $SU(3)$ shell 
model \cite{elliott}.
This provides the pseudo-$SU(3)$ irrep for $^{236}$U.
In a second step, the largest cluster is taken and its nucleons
filled into the Nilsson orbital at the {\it same deformation} as the united nucleus.
The number of nucleons in the active orbitals then determine the pseudo-$SU(3)$ irrep of the largest cluster. In a final step, the remaining nucleons, of the
lightest cluster, are filled {\it on top} of the largest cluster, which again
gives us the number of nucleons for the light cluster in the pseudo-oscillator.
In this manner, the number of nucleons in the active orbitals of the light plus of
the heavy cluster sum up to the nucleons in the active orbitals of the 
united nucleus.

This procedure is different to the pseudo-$SU(3)$ approach in 
\cite{cseh-scheid,cseh-algora}, where either different deformations were taken for the individual clusters and/or the light cluster was treated within the $SU(3)$ model
for light clusters. Here, we decided to take another approach, noting that
{\it all nucleons move in the same mean field 
with the same deformation} and that
the number of nucleons in the active orbitals are determined by the united nucleus.

In Table \ref{Tx1} 
several decay channels $^{236}$U $\rightarrow$
$X + Y$ are listed, whose forbiddenness was determined. The first column lists
the case number, to which Table \ref{Ty2} refers to.
This table shows the forbiddenness as obtained by our counting procedure.
The pseudo-$SU(3)$ irreps for each cluster, for the cluster irrep
$\left(\lambda_c,\mu_c\right)$ and the total irrep $\left(\lambda , \mu \right)$
are given in Table \ref{Ty2}. For the cluster irrep we take the largest one as 
suggested by our counting procedure. This irrep is then coupled with
$\left(n_c,0\right)$ and the relative irrep $\left( n_r,0\right)$ to the
total irrep of the united nucleus $^{236}$U, which is (54,0). This 
irrep is obtained by first determining the irreps for the proton and neutron parts
and then coupling them to the maximal irrep of the united nucleus. The number
of protons in the active levels is 46 and for the neutrons it is 82.

\vskip 0.5cm
\begin{table}
\begin{center}
\begin{tabular}{|c|c|c|c|}
\hline
{\rm No.} & 
{\rm Two}~{\rm cluster}~{\rm system} & United cluster & $n_c$ \\
\hline 
1 & $_{2}^{4}He_{2}$ $+$ $_{90}^{232}Th_{142}$ & 
$_{92}^{236}U_{144}$ & $0$\\
\hline 
2 & $_{10}^{20}Ne_{10}$ $+$ $_{82}^{216}Pb_{134}$ & 
$_{92}^{236}U_{144}$ & $4$\\
\hline 
3 & $_{10}^{24}Ne_{10}$ $+$ $_{82}^{212}Pb_{130}$ &  
$_{92}^{236}U_{144}$ & $4$\\
\hline 
4 & $_{10}^{26}Ne_{16}$ $+$ $_{82}^{210}Pb_{128}$ &  
$_{92}^{236}U_{144}$ & $2$\\
\hline 
5 & $_{12}^{28}Mg_{16}$ $+$ $_{80}^{208}Hg_{128}$ & 
$_{92}^{236}U_{144}$ & $4$\\
\hline 
6 & $_{12}^{30}Mg_{18}$ $+$ $_{80}^{206}Hg_{126}$ & 
$_{92}^{236}U_{144}$ & $4$\\
\hline 
7 & $_{14}^{32}Si_{18}$ $+$ $_{78}^{204}Pt_{126}$ & 
$_{92}^{236}U_{144}$ & $0$\\
\hline 
8 & $_{14}^{34}Si_{20}$ $+$ $_{78}^{202}Pt_{124}$ &  
$_{92}^{236}U_{144}$ & $6$\\
\hline 
9 & $_{22}^{40}Ti_{18}$ $+$ $_{70}^{296}Yb_{126}$ &  
$_{92}^{236}U_{144}$ & $16$\\
\hline 
10 & $_{36}^{66}Kr{}_{30}$ $+$ $_{56}^{170}Ba_{114}$ & 
$_{92}^{236}U_{144}$ & $20$\\
\hline 
11 & $_{22}^{66}Ti{}_{44}$ $+$ $_{70}^{170}Yb_{100}$ & 
$_{92}^{236}U_{144}$ & $32$\\
\hline 
12 & $_{50}^{128}Sn{}_{78}$ $+$ $_{42}^{108}Mo_{66}$ & 
$_{92}^{236}U_{144}$ & $28$\\
\hline 
13 & $_{50}^{132}Sn{}_{82}$ $+$ $_{42}^{104}Mo_{62}$ & 
$_{92}^{236}U_{144}$ & $36$\\
\hline 
\end{tabular}
\caption{
A series of 2-cluster systems, all belonging to the nucleus $_{92}^{236}U_{144}$
are enumerated (first column). In the last column the {\it forbiddenness} $n_c$
for these systems is evaluated.
\label{Tx1}
}
\end{center}
\end{table}

\vskip 0.5cm
\begin{table}
\begin{center}
\begin{tabular}{|c|c|c|c|c|c|}
\hline
{\rm No.} & 
$\left( \lambda_1,\mu_1\right)$ & $\left( \lambda_2,\mu_2\right)$ & 
$\left( \lambda_c,\mu_c\right)$ & $n_c$ & $\left(n_0,0\right)$  \\
\hline
1 & (0,0) & (48,4) & (48,4) & 0 & (10,0) \\ 
2 & (0,4) & (24,0) & (24,4) & 4 & (46,0) \\ 
3 & (0,2) & (16,2) & (16,4) & 4 & (54,0) \\ 
4 & (4,2) & (10,0) & (6,6) & 2 & (60,0) \\ 
5 & (4,0) & (10,6) & (8,6) & 4 & (64,0) \\ 
6 & (4,0) & (10,6) & (8,6) & 4 & (64,0) \\ 
7 & (8,0) & (12,8) & (4,16) & 0 & (66,0) \\ 
8 & (8,2) & (2,8) & (10,10) & 6 & (72,0) \\ 
9 & (10,0) & (22,0) & (32,0) & 16 & (70,0) \\ 
10 & (0,0) & (24,8) & (24,8) & 20 & (98,0) \\ 
11 & (16,4) & (36,0) & (52,4) & 32 & (102,0) \\ 
12 & (0,6) & (24,2) & (24,8) & 28 & (116,0) \\ 
13 & (12,0) & (8,0) & (20,0) & 36 & (118,0) \\
\hline
\end{tabular}
\caption{
List of pseudo-$SU(3)$ irreps. The first column refers to the number of the cluster system
as listed in Table \ref{Tx1}. The second and third
columns list the irrep for the first and second cluster, respectively. The fourth
column lists the cluster irreps, where we took always the largest one, according to
the findings in the text. The fifth column lists again the forbiddenness,
now written in terms of a pseudo-$SU(3)$ irrep, and the last column lists
the number of oscillation quanta which have to be added according to the
Wildermuth condition. 
\label{Ty2}
}
\end{center}
\end{table}

\begin{figure}
\begin{center}
\includegraphics[width=12cm,height=9cm]{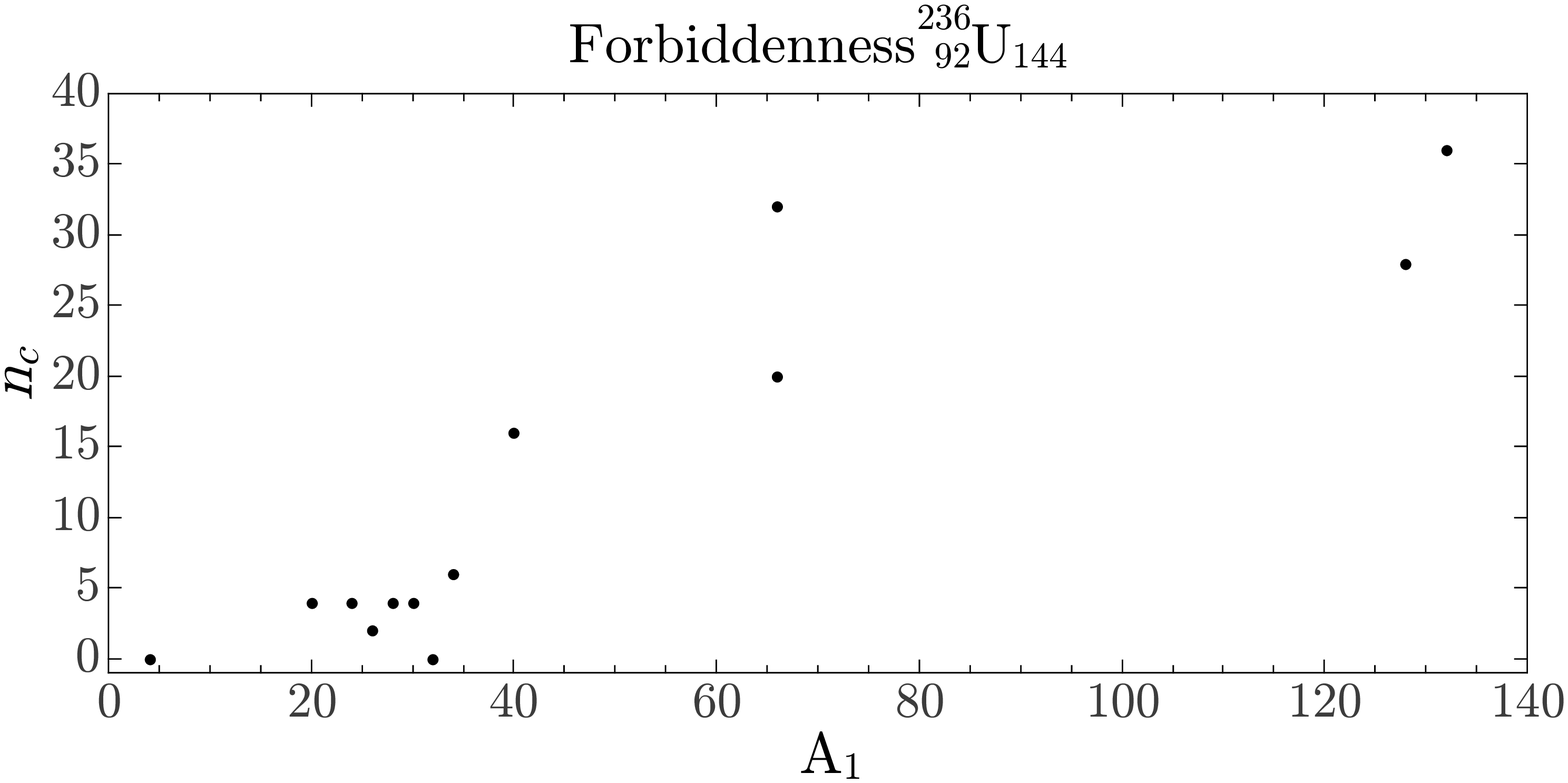}
\caption{Graphical representation of the results from Table \ref{Tx1}.
The forbiddenness is depicted versus the mass of the lightest cluster.
\label{graph1}
}
\end{center}
\end{figure}

\vskip 0.5cm
Fig. \ref{graph1} depicts the {\it forbiddeness} versus the nuclear mass of the 
lightest cluster. As a 
key feature we notice that when one cluster is light the
{\it forbiddenness} is zero or very low, while it
increases as a function of the mass of the lightest cluster.
It is in qualitative agreement with theories for cluster radioactivity \cite{sandulescu,poenaru,gupta},
i.e., that the probability to emit a cluster decreases with its mass,
which should be correlated with increasing {\it forbiddenness}.

\begin{table}[tbp]
\begin{center}
\begin{tabular}{|c|c|c|c|}
\hline
\textrm{No.} & \textrm{Two}~\textrm{cluster}~\textrm{system} & United cluster
& $n_{c}$ \\ \hline
1 & $_{2}^{4}He_{2}$ $+$ $_{96}^{248}Cm_{152}$ & $_{98}^{252}Cf_{154}$ & $0$
\\ \hline
2 & $_{8}^{16}O_{8}$ $+$ $_{90}^{236}Th_{146}$ & $_{98}^{252}Cf_{154}$ & $0$
\\ \hline
3 & $_{6}^{20}C_{14}$ $+$ $_{92}^{232}U_{140}$ & $_{98}^{252}Cf_{154}$ & $4$
\\ \hline
4 & $_{10}^{24}Ne_{14}$ $+$ $_{88}^{228}Ra_{140}$ & $_{98}^{252}Cf_{154}$ & $%
4$ \\ \hline
5 & $_{14}^{38}Si_{24}$ $+$ $_{84}^{214}Po_{130}$ & $_{98}^{252}Cf_{154}$ & $%
4$ \\ \hline
6 & $_{16}^{40}Sn_{24}$ $+$ $_{82}^{212}Pb_{130}$ & $_{98}^{252}Cf_{154}$ & $%
4$ \\ \hline
7 & $_{16}^{44}Sn_{28}$ $+$ $_{82}^{208}Pb_{126}$ & $_{98}^{252}Cf_{154}$ & $%
10$ \\ \hline
8 & $_{18}^{46}Ar_{28}$ $+$ $_{80}^{206}Hg_{126}$ & $_{98}^{252}Cf_{154}$ & $%
6$ \\ \hline
9 & $_{18}^{50}Ar_{32}$ $+$ $_{80}^{206}Hg_{122}$ & $_{98}^{252}Cf_{154}$ & $%
14$ \\ \hline
10 & $_{30}^{78}Zn_{48}$ $+$ $_{68}^{174}Er_{106}$ & $_{98}^{252}Cf_{154}$ & 
$18$ \\ \hline
11 & $_{30}^{80}Zn_{50}$ $+$ $_{68}^{172}Er_{104}$ & $_{98}^{252}Cf_{154}$ & 
$18$ \\ \hline
12 & $_{38}^{98}Sr_{60}$ $+_{60}^{152}Nd{}_{92}$ & $_{98}^{252}Cf_{154}$ & $%
24$ \\ \hline
13 & $_{38}^{100}Sr_{62}$ $+_{60}^{152}Nd{}_{92}$ & $_{98}^{252}Cf_{154}$ & $%
28$ \\ \hline
14 & $_{40}^{100}Zr_{60}$ $+_{58}^{152}Ce{}_{94}$ & $_{98}^{252}Cf_{154}$ & $%
28$ \\ \hline
15 & $_{40}^{102}Zr_{62}$ $+_{58}^{150}Ce{}_{92}$ & $_{98}^{252}Cf_{154}$ & $%
28$ \\ \hline
16 & $_{40}^{104}Zr_{64}$ $+_{58}^{148}Ce{}_{90}$ & $_{98}^{252}Cf_{154}$ & $%
28$ \\ \hline
17 & $_{42}^{104}Mo_{62}$ $+_{56}^{148}Ba{}_{92}$ & $_{98}^{252}Cf_{154}$ & $%
28$ \\ \hline
18 & $_{42}^{108}Mo_{66}$ $+_{56}^{144}Ba{}_{88}$ & $_{98}^{252}Cf_{154}$ & $%
28$ \\ \hline
19 & $_{44}^{110}Ru_{66}$ $+_{54}^{142}Xe_{88}$ & $_{98}^{252}Cf_{154}$ & $%
28 $ \\ \hline
20 & $_{44}^{112}Ru_{68}$ $+_{54}^{140}Xe_{86}$ & $_{98}^{252}Cf_{154}$ & $%
26 $ \\ \hline
21 & $_{44}^{114}Ru_{70}$ $+_{54}^{138}Xe_{84}$ & $_{98}^{252}Cf_{154}$ & $%
26 $ \\ \hline
22 & $_{46}^{116}Pd_{70}$ $+_{52}^{136}Te_{84}$ & $_{98}^{252}Cf_{154}$ & $%
26 $ \\ \hline
\end{tabular}%
\end{center}
\caption{ A series of 2-cluster systems, all belonging to the nucleus 
$_{98}^{252}$Cf$_{154}$, 
$(\protect\lambda,\protect\mu)=(56,10))$, are
enumerated (first column). In the last column the \textit{forbiddenness} $n_c
$ for these systems is evaluated. }
\label{Tx3}
\end{table}

\begin{table}[tbp]
\begin{center}
\qquad\ 
\begin{tabular}{|c|c|c|c|c|c|}
\hline
No. & $\left( \lambda _{1},\mu _{1}\right) $ & $\left( \lambda _{2},\mu
_{2}\right) $ & $\left( \lambda _{c},\mu _{c}\right) $ & $n_{c}$ & $\left( n_{0},0\right) $ \\ \hline
1 & (0,0) & (52,12) & (52,12) & 0 & (18,0) \\ 
2 & (4,0) & (52,6) & (52,8) & 0 & (32,0) \\ 
3 & (2,0) & (48,4) & (46,6) & 4 & (48,0) \\ 
4 & (0,2) & (42,6) & (42,8) & 4 & (54,0) \\ 
5 & (6,0) & (24,2) & (18,8) & 4 & (78,0) \\ 
6 & (10,0) & (16,2) & (14,8) & 4 & (82,0) \\ 
7 & (6,4) & (10,0) & (16,4) & 10 & (88,0) \\ 
8 & (6,6) & (10,6) & (16,12) & 6 & (90,0) \\ 
9 & (4,2) & (0,14) & (4,16) & 14 & (100,0) \\ 
10 & (12,0) & (28,12) & (16,24) & 18 & (120,0) \\ 
11 & (12,0) & (28,12) & (16,24) & 18 & (120,0) \\ 
12 & (8,8) & (32,4) & (36,14) & 24 & (126,0) \\ 
13 & (8,2) & (30,4) & (38,6) & 28 & (128,0) \\ 
14 & (8,2) & (30,4) & (38,6) & 28 & (128,0) \\ 
15 & (8,2) & (30,4) & (38,6) & 28 & (128,0) \\ 
16 & (16,2) & (30,0) & (38,6) & 28 & (128,0) \\ 
17 & (8,2) & (30,4) & (38,6) & 28 & (128,0) \\ 
18 & (16,2) & (30,0) & (34,8) & 28 & (128,0) \\ 
19 & (20,0) & (26,2) & (38,6) & 28 & (128,0) \\ 
20 & (24,2) & (20,4) & (36,10) & 26 & (128,0) \\ 
21 & (24,2) & (20,4) & (36,10) & 26 & (128,0) \\ 
22 & (22,6) & (18,2) & (36,10) & 26 & (128,0) \\ \hline
\end{tabular}%
\end{center}
\caption{ List of pseudo-$SU(3)$ irreps. The first column refers to the
number of the cluster system as listed in Table \protect\ref{Tx3}. The
second and third columns list the irrep for the first and second cluster,
respectively. The fourth column lists the cluster irreps, where we took
always the largest one, according to the findings in the text. The fifth
column lists again the forbiddenness, now written in terms of a pseudo-$SU(3)$ 
irrep, and the last column lists the number of oscillation quanta which
have to be Wildermuth condition. 
}
\label{Ty4}
\end{table}

\begin{figure}[tbp]
\begin{center}
\includegraphics[width=12cm,height=9cm]{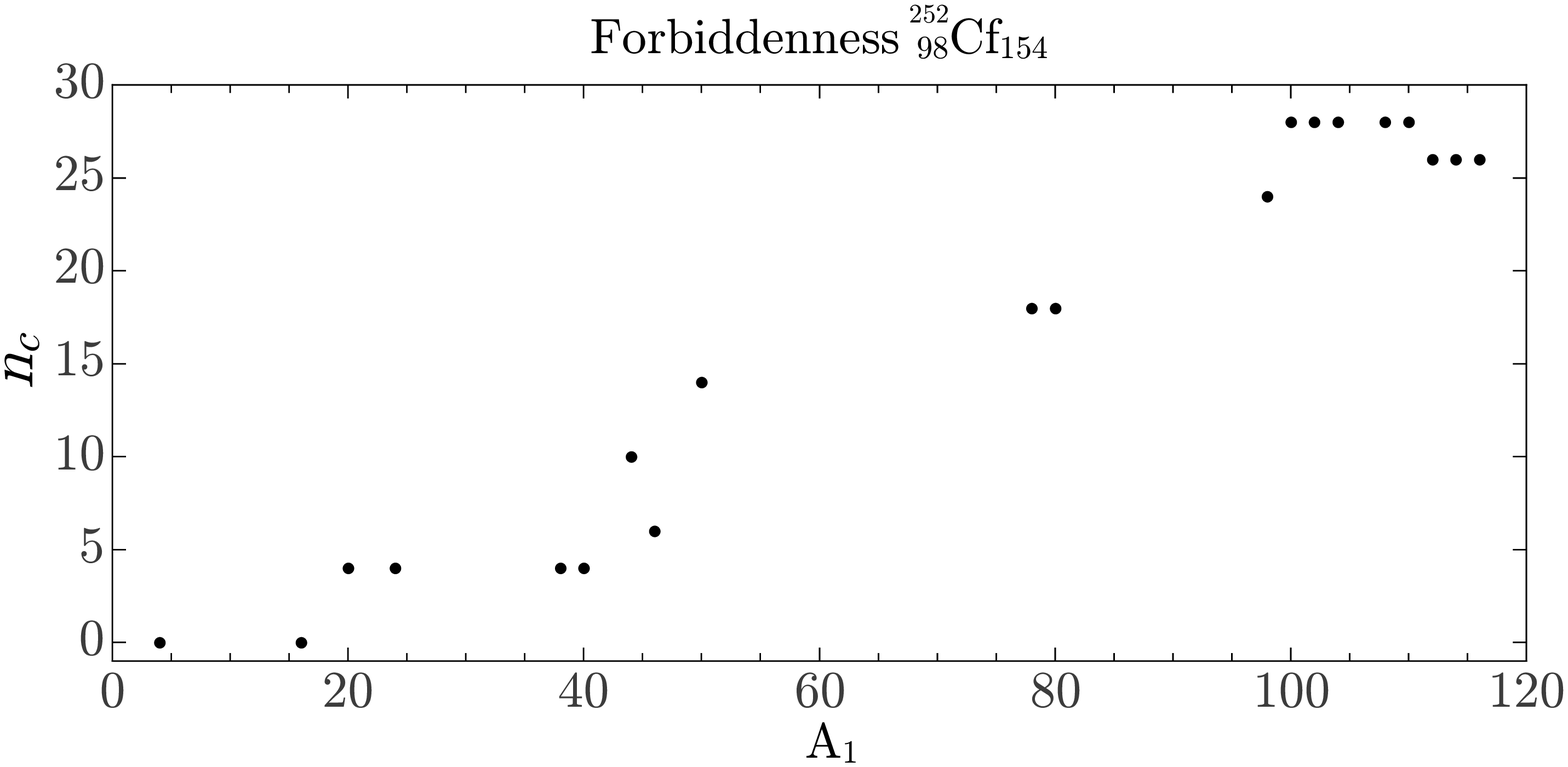}
\end{center}
\caption{Graphical representation of the results from Table \protect\ref{Tx3}%
. The forbiddenness is depicted versus the mass of the lightest cluster. }
\label{graph2}
\end{figure}

As a further example, we discuss the fission of $^{252}$Cf,
which is of great interest (please, consult the web page
of Ref. \cite{wolle} which contains all possible links to recent experimental activities related to $^{252}$Cf).
The procedure followed is the same as in $^{236}$U and the 
interpretation of the results is similar. In Table 
\ref{Tx3} a list of two-cluster systems
is given, with the same united $^{252}$Cf nucleus. In the 
last column the {\it forbiddenness} is listed. Again, the heavier
the lightest cluster is, the larger is $n_c$. 
For completeness, in Table
\ref{Ty4} the pseudo-$SU(3)$ irreps for each cluster system
is given. Finally, in Figure \ref{graph2}
the $n_c$ is plotted versus the mass number of the
lightest cluster. As in the former case, the {\it forbiddenness}
increases with the mass of the light cluster.

The introduction of {\it forbiddenness} will have another implication
for the determination of spectroscopic factors within algebraic models:
In \cite{spec-fac-serban,spec-fac-draayer} a very effective parametrization of 
the spectroscopic factor within the SACM was given. In \cite{spec-fac-draayer} this
parametrization was applied to nuclei in the p- and sd-shell with only few 
percent deviations to exact calculations within the $SU(3)$ 
model \cite{draayer-spec}. In \cite{spec-fac-serban} a simpler ansatz for the
spectroscopic factor was used to deduce the spectroscopic factors for 
fission products in $^{236}$U. There it was shown that the process of preformation, followed by the penetration
through the potential barrier, gives the same qualitative results
as in hydrodynamic models.
with following penetration through the potential barrier gives the same results
as for fission using hydrodynamic models, which is
stated differently in \cite{gupta}. At the low mass cluster side the model also
follows the linear scaling rule of Blendowske and 
Walliser \cite{waliser}, who suspected that this linear behavior has
to be changed for heavy clusters in order to reproduce the 
observed half-lives. In \cite{spec-fac-serban} the algebraic spectroscopic factor, within the SACM, was constructed and in a natural way reproduced the required deviation. Due to the explicit appearance of microscopic information, via the
pseudo-$SU(3)$ irreps, shell effects are automatically included.
The ansatz in \cite{spec-fac-serban} does not include
the $n_c$ dependence yet.

In \cite{spec-fac-serban,spec-fac-draayer} only dependencies 
of the spectroscopic factor on the number
of $\pi$-bosons, and the different $SU(3)$ irreps are given. With the introduction
of an additional dependence on the {\it forbiddenness}, $n_c$, we hope to improve the
description of the unification of the cluster radioactivity with the fission 
process, as intended in \cite{spec-fac-serban}.

\section{Conclusions}

We have not only presented a consistent procedure to calculate the {\it forbiddenness} of a two-cluster system but also determined which cluster irrep
has to be constructed in order to couple with the relative motion to the
ground state of the united nucleus. It has been demonstrated  that in general the
two clusters within the united nucleus have to be in 
an excited state and
not in their ground state. 
As the main results a
simple rule easy to implement emerged, namely in order to reduce 
the forbiddenness one has to take the largest, most compact 
$SU(3)$ 
cluster irrep (large $\left(2\lambda_c + \mu_c\right)$
but $\mu_c$ larger than $\lambda_c$).

The main idea of the procedure presented can be 
easily extended to the system of
more than two clusters, increasing the number of possibilities
of cluster systems to be analyzed.
 
Also  heavy cluster systems can be treated, 
where the pseudo-$SU(3)$ 
scheme has to be applied. As an example, we treated several decay 
channels of $^{236}$U and $^{252}$Cf
and discussed the importance of the 
results in connection with
the unification of cluster radioactivity with the fission process, i.e., that both
follow the same underlying physics of first pre-formation and
then penetration
through the barrier. 

There is also the possibility to improve the values for spectroscopic factors
\cite{spec-fac-serban,spec-fac-draayer},
introducing an additional dependence on the forbiddenness $n_c$.

\section*{Acknowledgment}
Th authors acknowledge financial help from DGAPA-PAPIIT (IN100315).


\vskip 1cm

\begin{thebibliography}{00}

\bibitem{smirnov} Yu. F. Smirnov and Yu. M. Tchuvil'sky, Phys. Lett. B 
{\bf 134} (1984), 25.

\bibitem{wildermuth-0} K.  Wildermuth and Th. Kanellopoulos  Nucl. Phys. {\bf 7} (1958) 150.

\bibitem{hecht} K. T. Hecht and A. Adler, Nucl. Phys. A {\bf 137} (1969), 129.

\bibitem{arima} A. Arima, M. Harvey and K. Shimizu, Phys. Lett. B {\bf 30}
(1969), 517.

\bibitem{wildermuth} K. Wildermuth and Y. C. Tang, {\it A Unified Theory of the
Nucleus}, (Academic Press, New York, 1977).

\bibitem{sacm1} J. Cseh, Phys. Lett. B {\bf 281} (1992), 173.

\bibitem{sacm2} J. Cseh and G. L\'evai, Ann. Phys. (N.Y.) {\bf 230} (1994), 165.

\bibitem{horiuchi} H. Horiuchi, J. Phys: Conf. Series
{\bf 569} (2014), 012001.

\bibitem{o16} C. Mahaux, Ann Rev Nucl Sci 23 
(1973), 193. 

\bibitem{RMF2008} J. Cseh, J. Darai, A. Algora, 
H. Y\'epez-Mart\ii nez and P. O. Hess, Rev. Mex. F\ii s.
S {\bf 54} (2008), 30.

\bibitem{sandulescu} A. Sandulescu, D. N. Poenaru and W. Greiner,
Sov. J. Part. Nucl. {\bf 11} (1980), 528.

\bibitem{poenaru} D.N.Poenaru and W.Greiner, in Handbook of Nuclear Properties, edited
by D.N.Poenaru and W.Greiner, (Clarendon Press, Oxford, 1996),
p.131.

\bibitem{gupta} R. Gupta and W. Greiner, Int. J. Mod. Phys. E {\bf 3} (1994), 335.

\bibitem{algora-2006} A. Algora, J. Cseh, J. Darai and P. O. Hess, Phys. Lett. B {\bf 639} (2006), 451.

\bibitem{cseh2010} J. Cseh et al., J. Phys.: Conf. Series
{\bf 239} (2010), 012006.

\bibitem{darai-2011} J. Darai et al., Phys. Rev. C {\bf 84} 
(2011), 024302.

\bibitem{darai-2012} J. Darai, J. Cseh and D. J. Jenkins, 
Phys. Rev. C {\bf 86} (2012), 064309.

\bibitem{spec-fac-serban} \c S. Mi\c sicu, P. O. Hess, 
Phys. Lett. B {\bf 595} (2004), 187.

\bibitem{cseh-2012} D. Lebhertz et al., Phys. Rev. C {\bf 85}
(2012), 034333. 

\bibitem{hoyle} F. Hoyle, The Astrophys. J. Suppl. Series {\bf 1}
(1954).

\bibitem{draayer-hoyle} A. C. Dreyfuss, K. D. Launey, T. Dytrych,
J. P. Draayer and C. Bahri, Phys. Lett. B {\bf 727} (2013), 511.

\bibitem{embedded1} P. Rochford and D. J. Rowe, Phys. Lett. B
{\bf 210} (1988), 5.

\bibitem{embedded2} D. J. Rowe, P. Rochford and J. Repka,
J. Math. Phys. {\bf 29} (1988), 572.

\bibitem{embedded3} M. Jarrio, J. L. Wood and D. J. Rowe, 
Nucl. Phys. A {\bf 528} (1991), 409.

\bibitem{hunyadi} P. O. Hess, M. Hunyadi, A. Algora and J. Cseh,
Eur. Phys. Jour. A15 (2002), 449. 

\bibitem{Yu} Yu.F. Smirnov and Yu M. Tchuvil'sky,
XXXIIth Conf. on Nuclear spectroscopy and nuclear
structure (Leningrad, 1982) p. 231.

\bibitem{Bader} R. Bader and P. Kramer, Physica A {\bf 114} (1982) 306.

\bibitem{cseh-forbiddeness} A. Algora and J. Cseh, J. Phys. G {\bf 22} (1996), L39.

\bibitem{cseh-tri-cluster} A. Algora, J. Cseh, J. Darai and P. O. Hess,
Phys. Lett. B {\bf 639} (2006), 451.

\bibitem{cseh-2009} W. Sciani et al., Phys. Rev. C {\bf 80}
(2009), 034319.

\bibitem{serban-heavy} P. O. Hess, \c S. Mi\c sicu, 
Phys. Rev. C {\bf 68} (2003), 064303.

\bibitem{hamermesh} M. Hamermesh, {\it Group Theory and its Application to
Physical problems}, (Dover Publications Inc, New York, 1989).

\bibitem{elliott} J. P. Elliott, Proc. R. Soc. A {\bf 245} (1958), 128
and 562.

\bibitem{pseudo-symplectic} O.Casta\~nos, P.O.Hess, P.Rocheford, J.P.Draayer,
Nucl. Phys. A {\bf 524} (1991), 469.

\bibitem{moller} P. M\"oller, J. R. Nix, W. D. Myers and W. J. Swiatecky,
Atomic Data Nucl. Data Tables {\bf 59} (1995), 185.

\bibitem{cseh-scheid} J. Cseh, R. K. Gupta and W. Scheid, Phys. Lett. 
B {\bf 299} (1993), 205.

\bibitem{cseh-algora} A. Algora and J. Cseh, J. Phys. G {\bf 22} (1996), L39.

\bibitem{wolle} H. J. Wollersheim, web-docs.gsi.de/~wolle

\bibitem{spec-fac-draayer} P. O. Hess, A. Algora, J. Cseh and J. P. Draayer, 
Phys. Rev. C {\bf 70} (2004), 051303(R).

\bibitem{draayer-spec} J. P. Draayer, Nucl. Phys. A {\bf 237} (1975), 157.

\bibitem{waliser} R. Blendowske and H. Walliser, 
Phys. Rev. Lett. {\bf 61} (1988), 1930.



\end{thebibliography}
\end{document}